\documentclass[aps,prl,showpacs,showkeys,twocolumn,superscriptaddress,floatfix,nobibnotes]{revtex4-2}
\usepackage[utf8]{inputenc}
\usepackage{verbatim}
\usepackage{amsmath}
\usepackage{mathtools}
\usepackage{gensymb}
\usepackage{amssymb}
\usepackage[pdftex]{graphicx}
\newcommand{\figurescale}{1}
\usepackage{soul}
\usepackage{xcolor}
\usepackage[normalem]{ulem}

\usepackage{soul}

\begin{document}

\title{Quantification of magnetic interactions in van der Waals heterostructures using Lorentz transmission electron microscopy and off-axis electron holography}

\affiliation{Ernst Ruska-Centre for Microscopy and Spectroscopy with Electrons, Forschungszentrum Jülich, Germany}
\affiliation{Peter Grünberg Institute, Forschungszentrum Jülich, Germany}
\affiliation{Department of Inorganic Chemistry, University of Chemistry and Technology Prague, Czech Republic}

\author{Joachim~Dahl~Thomsen}
\email{j.thomsen@fz-juelich.de}
\affiliation{Ernst Ruska-Centre for Microscopy and Spectroscopy with Electrons, Forschungszentrum Jülich, Germany}

\author{Qianqian~Lan}
\affiliation{Ernst Ruska-Centre for Microscopy and Spectroscopy with Electrons, Forschungszentrum Jülich, Germany}

\author{Nikolai~S.~Kiselev}
\affiliation{Peter Grünberg Institute, Forschungszentrum Jülich, Germany}

\author{Eva~Duft}
\affiliation{Ernst Ruska-Centre for Microscopy and Spectroscopy with Electrons, Forschungszentrum Jülich, Germany}

\author{Arslan~Rehmat}
\affiliation{Department of Inorganic Chemistry, University of Chemistry and Technology Prague, Czech Republic}

\author{Zdeněk~Sofer}
\affiliation{Department of Inorganic Chemistry, University of Chemistry and Technology Prague, Czech Republic}

\author{Rafal~E.~Dunin-Borkowski}
\affiliation{Ernst Ruska-Centre for Microscopy and Spectroscopy with Electrons, Forschungszentrum Jülich, Germany}

\date{\today}

\begin{abstract}
Magnetic van der Waals (vdW) materials are promising for memory and logic applications because of their highly tunable magnetic properties and compatibility with vdW heterostructure devices. However, coupling between magnetic textures in stacked layers is difficult to resolve in conventional plan-view measurements because the magnetic signal is integrated over the sample thickness. Here, these interactions are quantified in Fe$_3$GeTe$_2$ (FGT)/graphite/FGT heterostructures using cross-sectional Lorentz transmission electron microscopy and off-axis electron holography, enabling reconstruction of the local magnetic induction within and between the layers. Domain alignment weakens with increasing FGT separation, yielding a stray-field coupling length scale of $\lambda = 37 \pm 7$~nm for the cross-sectional geometry studied here, corresponding to the average separation at which domain misalignment first emerges. This length scale corresponds to an approximately 23\% reduction in the interlayer magnetic induction relative to bulk FGT. Surface effects result in a reduced magnetic induction compared to bulk FGT up to $\sim$100~nm from a surface. Comparisons of experimental data with model-based iterative reconstructions of the magnetization and micromagnetic simulations shows that the reduction in induction near surfaces is due to demagnetizing and stray fields. These results quantify the magnetic induction in stacked vdW magnets and guide the design of devices that require controllable coupling between magnetic textures.
\end{abstract}


\maketitle

\section{Introduction}

Magnetic van der Waals (vdW) materials have attracted significant attention following recent reports of intrinsic magnetism in bilayer Cr$_2$Ge$_2$Te$_6$ \cite{gong2017discovery}, monolayer CrI$_3$ \cite{huang2017layer}, and monolayer Fe$_3$GeTe$_2$ (FGT) \cite{deng2018gate}. Their magnetic properties can be tuned by layer thickness \cite{gong2017discovery, li2018patterning}, electrical stimuli \cite{deng2018gate, verzhbitskiy2020controlling, xing2017electric, zhuo2021manipulating, han2025electric}, pressure \cite{sun2018effects}, and strain \cite{vsivskins2022nanomechanical}, making them promising candidates for spintronic memory and logic devices, as well as for quantum information applications \cite{wang2022magnetic, mak2019probing, psaroudaki2021skyrmion}.

\begin{figure*}[]
	\scalebox{\figurescale}{\includegraphics[width=1\linewidth]{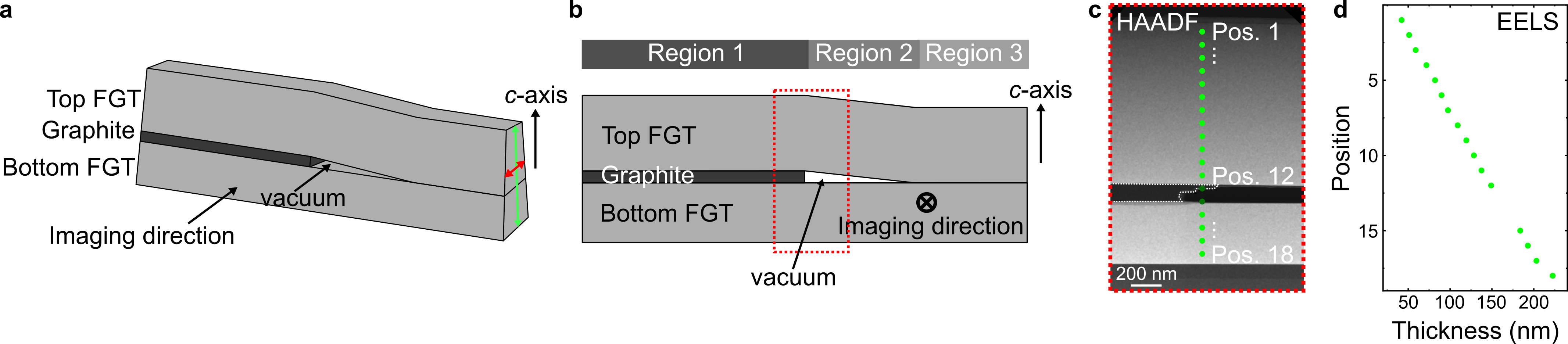}}
	\caption{\label{fig:schematic}
        \textbf{Sample overview.}
        \textbf{(a, b)} Schematic of HS1 shown in (a) an oblique view and (b) the cross-sectional TEM viewing orientation. The vertical black arrows indicate the orientation of the crystallographic \textit{c}-axis (the [0001] direction) of FGT and graphite. The red and green arrows indicate the lamella thickness and the FGT/graphite/FGT heterostructure thickness, respectively.
        \textbf{(c)} High-angle annular dark-field scanning transmission electron microscopy (HAADF-STEM) image of the region marked by the red rectangle in (b). The green dots indicate the approximate positions at which EEL spectra were acquired. EEL spectra were not acquired from the vacuum spacer at positions 13 and 14 (denoted by dark green dots).
        \textbf{(d)} Lamella thickness determined at the positions of the green dots in (c); see \textit{Methods} for details.        
            }
\end{figure*}

Another advantage of vdW magnets is that they can be assembled into heterostructures without lattice-matching constraints \cite{castellanos2022van}. This enables studies of interlayer coupling between vdW crystals, including magnetic interactions between distinct vdW magnets \cite{wu2022van}, twist-angle effects in homobilayers \cite{song2021direct, xu2022coexisting}, and proximity phenomena such as proximity-induced spin-orbit coupling \cite{wu2020neel}. In addition, magnetic vdW heterostructures are promising for spintronic applications. Relevant device concepts include vdW-based spin-orbit torque random-access memory \cite{shao2021roadmap}, magnetic tunnel junctions \cite{jia2025spintronic}, and spin valves \cite{wang2018tunneling}. In such devices, ferromagnetic layers are separated by non-magnetic spacers. Understanding the interactions between ferromagnetic layers, the resulting magnetic domain structures, and the effects of interfaces and surfaces on the local magnetization and magnetic induction is therefore important for the development of new spintronic applications.

However, in vdW heterostructures imaged using conventional magnetic imaging methods in the plan-view geometry, it is difficult to determine which magnetic layer gives rise to a given signal, because the measured signal is integrated through the thickness of the heterostructure. 

To address this challenge, we prepare cross-sectional lamellae of FGT/graphite/FGT heterostructures, in which the vertical separation between the FGT layers is controlled by the thickness of a graphite spacer. Cross-sectional magnetic imaging has previously been used successfully to clarify magnetic domain configurations in vdW magnets such as Fe$_5$GeTe$_2$ \cite{ni2026mesoscale}, Cr$_2$Ge$_2$Te$6$ \cite{han2019topological}, and FGT \cite{ding2022tuning}. FGT is a metallic vdW ferromagnet with a bulk Curie temperature ($T{\mathrm{c}}$) of $\sim$220~K and strong perpendicular magnetic anisotropy \cite{deiseroth2006fe3gete2, chen2013magnetic}. We investigate heterostructures composed of bulk-like FGT layers (thickness $\gtrsim$50~nm along the crystallographic \textit{c}-axis) separated by spacers ranging from monolayer graphene to 110~nm-thick graphite.

Using Lorentz transmission electron microscopy (LTEM) and off-axis electron holography (OAEH), we visualize the magnetic domain structure in each FGT layer. In addition, OAEH measures the projection of the in-plane components of the three-dimensional magnetic induction, $\boldsymbol{B}=\mu_{0}(\boldsymbol{M}+\boldsymbol{H})$, integrated along the electron-beam direction, where $\mu_{0}$ is the vacuum permeability, $\boldsymbol{M}$ is the magnetization, and $\boldsymbol{H}$ is the magnetic field intensity. In the absence of an applied magnetic field, $\boldsymbol{H}$ corresponds to the demagnetizing field inside the specimen and the stray field outside the specimen.

The cross-sectional geometry enables direct visualization of stray-field coupling and domain alignment as a function of FGT layer separation. Domain alignment decreases with increasing separation, although interactions persist even at a spacing of 110~nm. From one heterostructure, we determine a characteristic stray-field coupling length of $\lambda = 37 \pm 7$~nm, corresponding to the average FGT separation at which domain misalignments are first observed. OAEH further reveals that this separation corresponds to an approximately 23\% reduction in the interlayer stray field relative to the expected bulk FGT magnetic induction. In addition, the projected in-plane magnetic induction is reduced near the FGT surfaces over distances of up to $\sim$100~nm compared with the bulk value. We further discuss how the lamella geometry is expected to influence $\lambda$. Finally, comparison of the experimental data with model-based iterative reconstructions of the magnetization and micromagnetic simulations enables separating the contributions of $\boldsymbol{M}$ and $\boldsymbol{H}$, indicating that the reduced projected in-plane magnetic induction near the FGT surfaces arises from demagnetizing and stray fields.

Beyond their fundamental interest, these results provide design guidance and insight into surface effects on the local magnetic induction in devices that require controllable coupling between magnetic textures in vertically stacked vdW layers. More broadly, this cross-sectional imaging strategy enables high-resolution quantification of the magnetic induction and is applicable to a wide range of magnetic vdW heterostructures, including systems exhibiting proximity-induced phenomena arising from strong spin-orbit coupling, as well as systems with superconductor/ferromagnet interfaces.

\begin{figure*}[]
	\scalebox{\figurescale}{\includegraphics[width=1\linewidth]{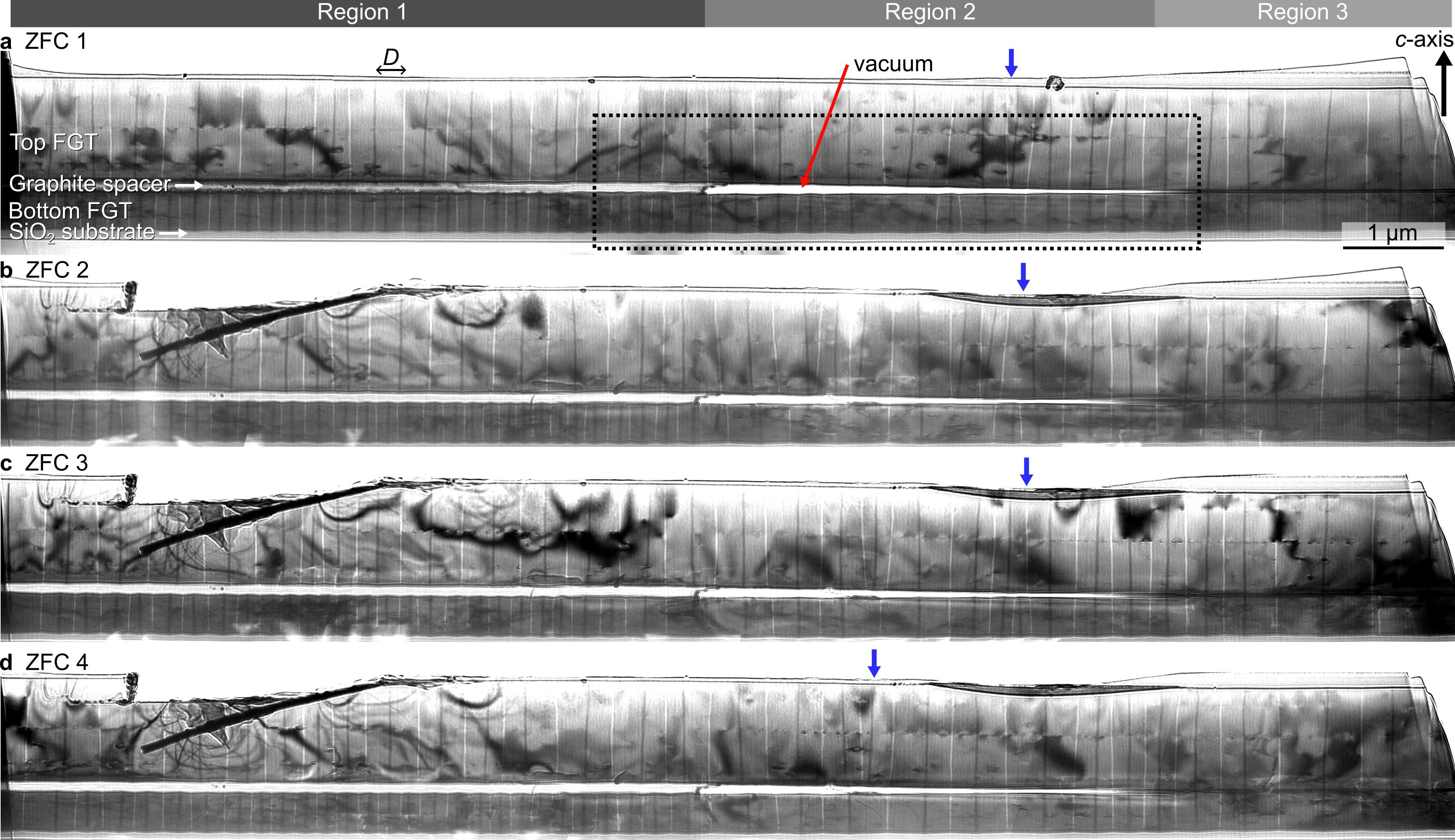}}
	\caption{\label{fig:overview}
        \textbf{Magnetic structure of HS1 after ZFC.}
        \textbf{(a--d)} Lorentz TEM images acquired at 95~K with a defocus of 0.3~mm. Each image was acquired following an identical ZFC procedure. The blue arrows indicate the positions of the first misalignments between domain walls in the top and bottom FGT layers. Some structural damage is visible in the upper part of the lamella in panels (b--d), which developed after several temperature cycles between 95 and 300~K. The domain width, $D$, is indicated in (a).        
        }
\end{figure*}

\section{Results and Discussion}

\subsection{Magnetic domain structures of FGT-graphite-FGT heterostructures}

The magnetic domain structure and stray-field coupling length scales in  FGT/graphite/FGT heterostructures were first determined using LTEM. In total, four vdW heterostructure lamellae were fabricated. The heterostructures were assembled from mechanically exfoliated FGT and graphite crystals, and the lamellae were subsequently prepared by focused ion-beam (FIB) milling; see \textit{Methods} for details. The main text focuses on results from heterostructure 1 (HS1) (see Fig.~S1 for additional characterization of HS1), while results from HS2--HS4 are summarized below and presented in the Supplementary Information, Fig.~S2--S4, respectively.

Figure~\ref{fig:schematic} shows a schematic of HS1 in two viewing orientations. Overview scanning electron microscopy (SEM) and optical microscopy images are provided in Fig.~S1. These images suggest that the lamella thickness along the electron-beam direction varies smoothly across the heterostructure, as illustrated schematically in Fig.~\ref{fig:schematic}. We subsequently measure the lamella thickness using electron energy-loss spectroscopy (EELS), as shown in Fig.~\ref{fig:schematic}(c, d), and find that the top and bottom FGT layers have thicknesses of approximately 40--150~nm and 180--220~nm in the electron beam direction, respectively. The inelastic mean free path used for the EELS thickness measurements was calibrated using spectra acquired from exfoliated FGT flakes with thicknesses independently measured by atomic force microscopy; see Fig.~S5 and \textit{Methods} for details. The average thickness measured immediately above and below the spacer is 166~$\pm$~17~nm, where the uncertainty represents the standard error of the mean, providing an estimate of the lamella thickness at the spacer position.
The OAEH measurements presented in the next section are consistent with a smoothly varying lamella thickness. The graphite spacer is 110~nm thick along the crystallographic \textit{c}-direction and occupies only the left half of the lamella (region~1, as indicated in Fig.~\ref{fig:schematic}(b)). In the central part of the heterostructure (region~2), the two FGT layers are separated by a vacuum gap that tapers laterally until the layers come into contact on the right side (region~3). Thus, region~2 contains a vacuum gap, whereas in region~3 the two FGT layers are in direct contact.

Throughout this study, the thickness along the imaging (electron-beam) direction is referred to as the lamella thickness, whereas the thickness of the FGT crystals along the crystallographic \textit{c}-direction is referred to as the FGT thickness. These two thickness directions are indicated by the red and green arrows in Fig.~\ref{fig:schematic}(a), respectively. Plan-view imaging of as-exfoliated FGT flakes is also considered below. In this geometry, the FGT-thickness direction is approximately parallel to the imaging direction, apart from the sample tilt required to obtain domain wall contrast.

Figure~\ref{fig:overview}(a--d) shows LTEM images of HS1 acquired at 95~K. The vertical dark and bright lines correspond to magnetic domain walls, while broader bands of dark contrast arise from diffraction contrast. Each image in Fig.~\ref{fig:overview} was obtained after zero-field cooling (ZFC) from room temperature, above the $T_{\mathrm{c}}$ of FGT, to 95~K inside the TEM. Repeating this thermal cycle between acquisitions enables comparison of domain wall configurations obtained after independent ZFC cycles. The domain wall configuration varies between cooling cycles, indicating that the domain walls are not pinned to specific locations.

\begin{figure}[t]
	\scalebox{\figurescale}{\includegraphics[width=1\linewidth]{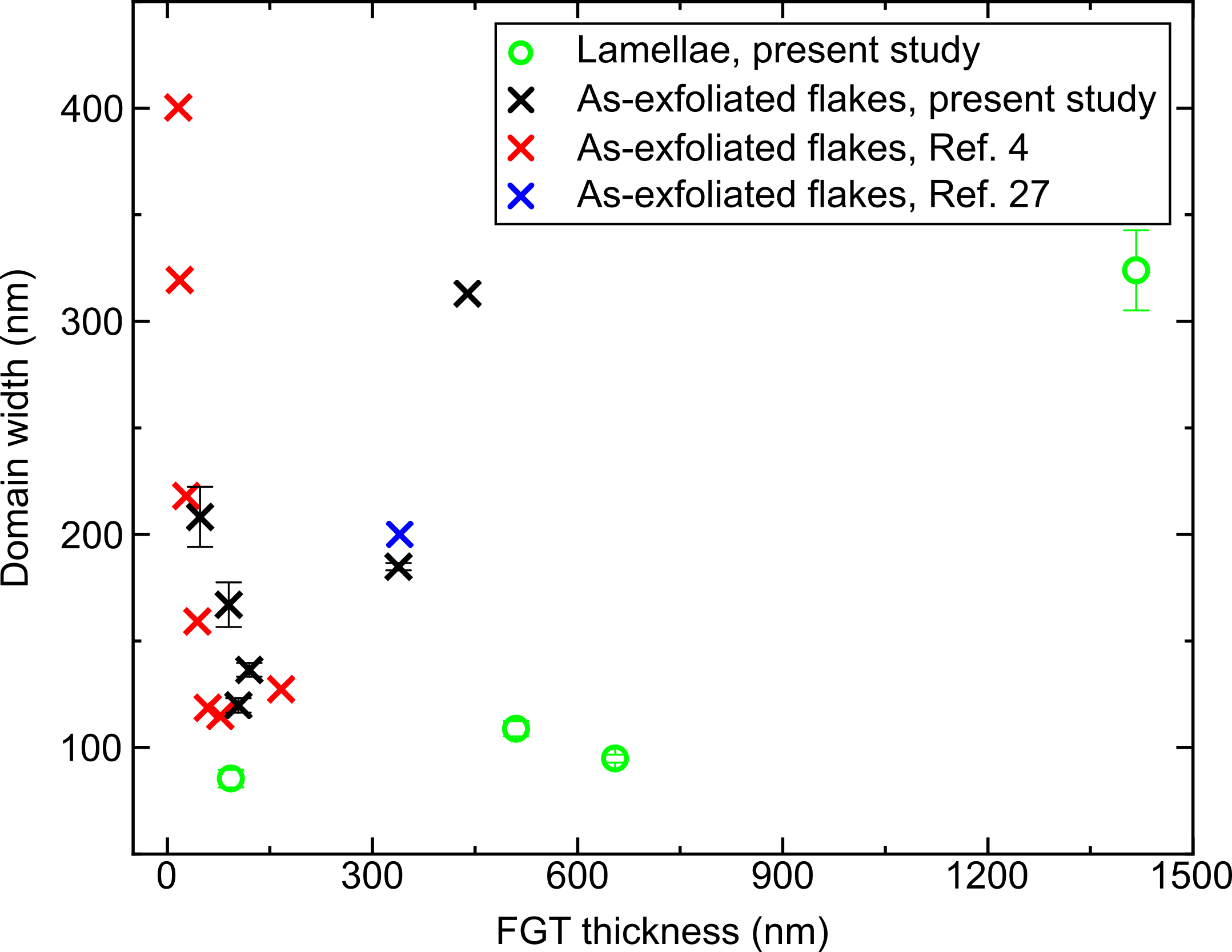}}
	\caption{\label{fig:lamella-vs-flake}
        \textbf{Domain width as a function of FGT thickness for FIB-prepared lamellae and as-exfoliated flakes.}
        The error bars represent the standard error of the mean. The plot includes data for as-exfoliated flakes extracted from Refs.~\cite{li2018patterning} and \cite{fei2018two} (red and blue crosses, respectively). The data point corresponding to the largest FGT thickness is obtained from region~3 of HS1. The domain widths for HS2--HS4 are measured from Figs.~S2(c), S3(b), and the right side of S4(c), respectively. These FGT thicknesses thus correspond to the combined thickness of two stacked FGT layers. The thicknesses of the as-exfoliated flakes were measured by atomic force microscopy, and the corresponding LTEM images are shown in Figs.~S7 and S13.
        }
\end{figure}

In regions~2 and~3, magnetic domains in the top and bottom FGT layers remain aligned at small layer separations and become misaligned once the separation is large enough to weaken the stray-field-mediated coupling. The first misalignments between domains in the two layers are marked by blue arrows in Fig.~\ref{fig:overview}. The FGT layer separations at the onset of misalignment are measured from in-focus images of the specimen at the positions indicated by the blue arrows; see Fig.~S6 for details. The mean FGT layer separation is $37 \pm 7$~nm, where the uncertainty is the standard error of the mean from four measurements. We use this value to define the stray-field coupling length scale for this geometry, $\lambda = 37$~nm. Strong coupling occurs for separations shorter than $\lambda$, whereas misalignment between the magnetic domains can occur for separations larger than $\lambda$.

Table~1 summarizes the average domain width, $D$ (indicated in Fig.~\ref{fig:overview}(a)), in each FGT layer and region of HS1. A quantitative comparison of the domain widths between the two layers is challenging because the lamella thickness is smaller in the top FGT layer than in the bottom layer. Moreover, Kittel's model ($D \propto \sqrt{W}$, where $W$ is the FGT thickness) \cite{kittel1946theory} is not directly applicable here, as it assumes a thin film with effectively infinite lateral extent. Nevertheless, $D$ is smaller in region~1, where the layers are separated, than in region~3, where they are in contact and can be approximated as a single, effectively thicker film. This qualitative trend is consistent with the predictions of Kittel's model.

The domain structure in region~1 also differs markedly from that in region~2, with the bottom FGT layer exhibiting nearly twice the domain periodicity of the top layer. This may simply relate to the spacer distance and the expected difference in $D$ as predicted by Kittel's model. In addition, EELS measurements show that the lamella thickness within the bottom FGT layer is approximately 15~nm larger in region~1 than in region~2 (Fig.~S1(h, i)), which may contribute to the abrupt change in domain structure. Finally, micromagnetic simulations (Fig.~\ref{fig:simulation}(e--g)) show that interlayer domain alignment can stabilize larger domains, which may also contribute to the differences in $D$ observed between the two FGT layers in region~1.

\subsubsection{Effect of FIB patterning on magnetic structure}

To investigate the effect of FIB patterning of FGT flakes into lamellae, we compare the domain width, $D$, after ZFC in FGT lamellae and as-exfoliated FGT flakes. Figure~\ref{fig:lamella-vs-flake} shows $D$ as a function of FGT thickness for the lamellae studied here and for as-exfoliated plan-view flakes. Above an FGT thickness of about 100~nm, the domain width increases with increasing FGT thickness for both lamellae and flakes. However, for the same FGT thickness, the domain widths in the lamellae are smaller than those in the exfoliated flakes. Thus, FIB processing results in a reduction in $D$.

\begin{table*}[]
\centering
\caption{Average domain widths in Regions~1--3, obtained by measuring all domain widths in the four LTEM images shown in Fig.~\ref{fig:overview}. The error bars represent the standard error of the mean. The number of domains, $n$, included in each average is given in parentheses.}
\label{tab:1}
\begin{tabular}{p{2.1cm} p{3.2cm} p{3.2cm} p{3.2cm} p{0cm}} \hline
\centering Region &  \centering 1 &  \centering 2 & \centering 3 & \\ \hline 
\centering Top FGT & \centering $249\pm7$ nm ($n$=104) & \centering $236\pm11$ nm ($n$=74) & \centering $324\pm19$ nm ($n$=28) & \\  
\centering Bottom FGT & \centering $119\pm5$ nm ($n$=210) & \centering $242\pm12$ nm ($n$=68) & \centering $324\pm19$ nm ($n$=28) &\\ \hline 
\end{tabular}
\end{table*}

In the following, we discuss the possible microscopic origins of the reduction in $D$, which may arise from (1) a decrease in the effective magnetic anisotropy, (2) additional stray-field energy associated with the specimen geometry, or (3) FIB-induced surface damage.

(1) FIB patterning has previously been reported to reduce the effective magnetic anisotropy in FGT \cite{li2018patterning}. In our case, shaping FGT into a lamella modifies the shape anisotropy, which in this geometry favors in-plane magnetization. The shape anisotropy does not overcome the magnetocrystalline anisotropy, since alternating out-of-plane domains are still observed; however, the effective magnetic anisotropy may nevertheless be reduced. A reduction in the effective magnetic anisotropy would in turn lead to smaller domain widths.

(2) Another possibility is a change in the stray-field energy. In typical exfoliated flakes with lateral dimensions on the order of several $\mu$m, the contribution of the side surfaces to the stray-field energy is negligible. In contrast, the lamellae studied here have comparable lamella and FGT thicknesses. The reduced domain width could therefore arise from additional stray-field energy associated with the side surfaces, which can be reduced by the formation of additional domain walls.

(3) FIB processing results in the formation of an amorphous surface layer \cite{kato2004reducing}, whose extent and magnetic effects are material dependent. In ferromagnets, FIB processing has been shown to produce a magnetically dead surface layer \cite{park2002local}. Conversely, in chiral magnets such as FeGe and Co$_8$Zn$_8$Mn$_4$, comparisons between experimental data and micromagnetic simulations suggest that the Dzyaloshinskii--Moriya interaction (DMI) reduces to zero in the surface layer while leaving the saturation magnetization and exchange stiffness intact \cite{zheng2023hopfion, zhu2026light}. Such surface modifications therefore do not necessarily affect the domain structure or magnetic properties in the bulk. Oxidation also leads to the formation of an amorphous surface layer in FGT, which has been reported to be antiferromagnetic and thereby reduces the effective ferromagnetic thickness of the crystal \cite{gweon2021exchange, li2023visualizing}. In the present specimens, FIB-induced surface damage could combine with oxidation to form a magnetically altered surface layer. Surface damage is minimized as much as possible by performing the final polishing steps at lower ion-beam acceleration voltages; see \textit{Methods} for details. Because the top of HS1 has a thickness as low as 40~nm (Fig.~1), with magnetic domains visible throughout this region (Fig.~2), any magnetically altered surface layer must be less than 20~nm thick.

Regardless of the microscopic origin, these results show that FIB patterning alters the domain structure in FGT. This effect should therefore be considered when comparing lamella-based measurements with measurements of as-exfoliated flakes, although it does not affect comparisons among the heterostructures studied here.

\begin{table*}[]
\centering
\caption{Lamella geometry and estimated values of $\lambda$. The listed lamella thicknesses are average values obtained from EELS measurements, except for HS1, for which we use the average of the EELS measurements acquired immediately above and below the graphite spacer. The listed FGT thickness is the combined thickness of the top and bottom FGT layers. For HS1, the approximate domain width in Region~2 is used. The remaining domain widths are taken from Fig.~3; see the figure caption for details of how they were determined. The determination of $\lambda$ for HS2 and HS3 is described in the main text. All uncertainties represent the standard error of the mean.}
\label{tab:2}
\vspace{0.1cm}
\begin{tabular}{p{2.5cm} p{3.9cm} p{3.5cm} p{3.5cm} p{2cm} p{0cm}} \hline
\centering Heterostructure &  \centering Lamella thickness (nm) &  \centering FGT thickness (nm)  & \centering Domain width (nm) & $\lambda$ (nm) & \\ \hline 
\centering 1 & \centering 166~$\pm$~17 & \centering 1417 & \centering $\sim$240$~\pm$~10 & 37~$\pm$~7 & \\  
\centering 2 & \centering 390~$\pm$~1 & \centering 655 & \centering 95~$\pm$~7 & $>$0.3 &\\ 
\centering 3 & \centering 279~$\pm$~0.5 & \centering 510 & \centering 109~$\pm$~17 & $>$30 &\\ 
\centering 4 & \centering 130~$\pm$~1 & \centering 93 & \centering 85~$\pm$~15 & $\sim$60 &\\ \hline 
\end{tabular}
\end{table*}

\subsubsection{The role of lamella geometry in magnetic interactions}

In the following, we discuss the expected role of lamella geometry by considering three factors: (1) lamella thickness, (2) FGT thickness, and (3) domain width. As a first approximation, we assume that each parameter can be varied independently of the others. The effects of varying (1) and (2) can be understood in terms of magnetic surface charges arising from the discontinuity in magnetization at the FGT surfaces. A schematic illustration is provided in Fig.~S8.

(1) Increasing the lamella thickness is expected to strengthen the coupling and therefore increase $\lambda$, because the interfacial area between the coupled FGT layers increases.

(2) Increasing the FGT thickness is expected to weaken the coupling because of the reduced interaction between magnetic surface charges at the outer surfaces of the FGT layers (see Fig.~S8). This would result in a decrease in $\lambda$.

(3) Increasing the domain width is expected to increase the stray field outside the sample and therefore enhance the dipole-dipole coupling between the FGT layers and increase $\lambda$.

However, beyond these intuitive trends, the domain width itself is expected to depend on both the lamella thickness and the FGT thickness, as illustrated for the latter in Fig.~3. In FGT, the domain size exhibits a non-monotonic dependence on thickness: it initially decreases with increasing thickness, reaches a minimum, and then increases, likely following Kittel scaling ($D \propto \sqrt{W}$) at larger thicknesses \cite{li2018patterning, chakraborty2022magnetic}. Because these parameters are interdependent, $\lambda$ is unlikely to exhibit a simple scaling with any of the three individually.

The lamella dimensions, domain widths, and measured values of $\lambda$ for HS1--HS4 are summarized in Table~2. The spacers in HS2 and HS3 are monolayer graphene and graphite with a spatially varying thickness, respectively. In HS3, the graphite spacer reaches a maximum thickness of 30~nm. In both heterostructures, the magnetic domains in the top and bottom FGT layers remain fully aligned. We therefore estimate that $\lambda > 0.3$~nm, corresponding to the thickness of monolayer graphene for HS2, and $ \lambda > 30$~nm for HS3.

Comparing HS1 and HS4, both of which yield an estimate of $\lambda$, we find that HS4 has a much smaller FGT thickness than HS1. This would be expected to increase $\lambda$, consistent with the experimental observations. However, the domain widths in HS4 are also smaller than those in HS1, which would tend to decrease $\lambda$. These data therefore suggest that, when comparing these two heterostructures, the effect of the smaller FGT thickness dominates, resulting in a larger $\lambda$ for HS4.

We emphasize that $\lambda$ for HS4 was determined from a single ZFC process and therefore likely carries a larger uncertainty than our estimate for HS1. In addition, a systematic study combining measurements from several specimens with micromagnetic simulations would be required to obtain a quantitative understanding of the influence of specimen geometry on $\lambda$. Finally, the estimates for $\lambda$ from HS1 and HS4 come from a FGT/vacuum/FGT region. To clarify whether $\lambda$ is affected by the spacer medium, a heterostructure with spatially varying graphite thickness in a suitable thickness range would need to be investigated.

\begin{figure*}[]
	\scalebox{\figurescale}{\includegraphics[width=1\linewidth]{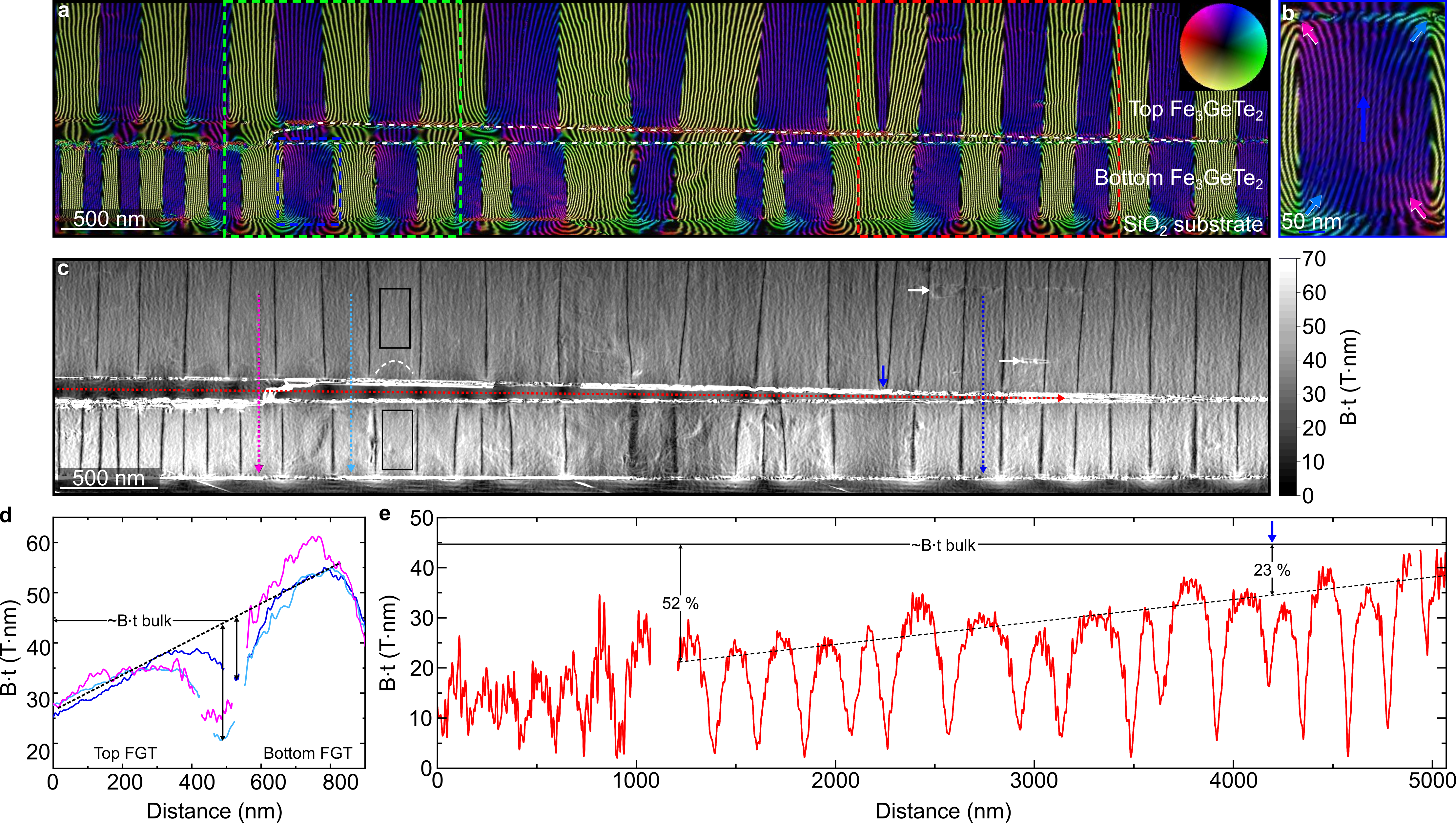}}
	\caption{\label{fig:Bfield}
        \textbf{Magnetic field distribution.}
        \textbf{(a)} Magnetic contour map acquired at 95~K by OAEH. The contour spacing is $2\pi/7$~rad. A higher density of contour lines corresponds to a larger magnetic field.
        \textbf{(b)} Magnetic contour map from the blue rectangle in (a). Arrows indicate the local direction of the magnetic induction near the domain corners, with color denoting the direction.
        \textbf{(c)} In-plane projected magnetic induction, $|\mathbf{B}_{\perp}|\,t$. The average values within the upper and lower black rectangles are 30.9~T\,nm and 50.6~T\,nm, respectively. The striped white arc highlights a typical near-surface region within a domain where $|\mathbf{B}_{\perp}|\,t$ is reduced. The horizontal features indicated by the white arrows arise from diffraction contrast, which is also present in the LTEM images in Fig.~\ref{fig:overview}. The blue arrow indicates the first instance of domain misalignment in region~2.
        \textbf{(d)} Line profiles of $|\mathbf{B}_{\perp}|\,t$ along the pink line (region~1) and the light- and dark-blue lines (region~2). Edge artifacts produce artificially large values at the FGT boundaries; these segments have therefore been removed, resulting in gaps near the vacuum regions. The dotted black line extrapolates the trend of $|\mathbf{B}_{\perp}|\,t$ from the interior of the FGT across the vacuum spacer, while the two vertical arrows indicate approximately the center of the vacuum spacer for the two region~2 profiles. The reference value of $|\mathbf{B}_{\perp}|\,t_{\mathrm{bulk}}$ for the vacuum spacer is taken from the extrapolated dotted line between these positions.
        \textbf{(e)} Line profile of $|\mathbf{B}_{\perp}|\,t$ along the red line in (c). The black dashed line is a guide to the eye showing the increase in $|\mathbf{B}_{\perp}|\,t$ as the interlayer separation decreases. This guide is used to estimate the reduction in $|\mathbf{B}_{\perp}|\,t$ within the vacuum spacer relative to the extrapolated interior FGT value.
		} 
\end{figure*}

\subsection{Visualization of the internal and stray magnetic field}

OAEH was then used to quantify the magnetic fields within and between the FGT layers. In all datasets, the electrostatic contribution to the phase shift was removed by subtracting the phase measured at room temperature, above the $T_{\mathrm{c}}$ of FGT; see \textit{Methods} for details.

Figure~\ref{fig:Bfield}(a) shows a magnetic contour map of the region indicated by the black dotted rectangle in Fig.~\ref{fig:overview}(a). The map directly visualizes the local magnetic field lines within and between the FGT layers, as well as the stray fields in the SiO$_2$ substrate. Closer inspection of a magnetic domain in Fig.~\ref{fig:Bfield}(b) shows that the magnetic field lines are aligned with the crystallographic \textit{c}-direction in the central region of the domain, while they bend away from the \textit{c}-direction near the domain corners. The decreased density of lines due to this bending near the surfaces indicates a reduced projected in-plane magnetic induction. 

Next, we quantify the magnetic field distribution in the heterostructure. For a specimen of uniform composition, the projected in-plane magnetic induction is given by \cite{dunin2019electron}

\begin{equation}\label{eq:B}
    \boldsymbol{B}_{\perp}(x,y) = -\left( \frac{\hbar}{et} \right) \nabla \phi(x,y) ,
\end{equation}

where $\phi$ is the magnetic phase shift of the electron wave, $e$ is the electron charge, $t$ is the lamella thickness, and $\hbar$ is the reduced Planck constant.

Examples of holograms 
used to make Fig.~\ref{fig:Bfield}(a) are shown in Fig.~S9. Because the lamella does not have a uniform thickness, we use the magnitude of the thickness-integrated projected in-plane magnetic induction, $|\boldsymbol{B}_{\perp}(x,y)|\,t$, shown in Fig.~\ref{fig:Bfield}(c), for the analysis. Bright bands separated by narrow domain walls are observed, consistent with alternating magnetic domains in FGT.

The magnitude $|\boldsymbol{B}_{\perp}(x,y)|\,t$ is larger in the bottom FGT layer than in the top layer, consistent with the greater lamella thickness in the lower part of the specimen. Figure~\ref{fig:Bfield}(d) shows line profiles of $|\boldsymbol{B}_{\perp}(x,y)|\,t$ extracted along the vertical lines in Fig.~\ref{fig:Bfield}(c). Within the interior of the FGT layers, $|\boldsymbol{B}_{\perp}(x,y)|\,t$ varies approximately linearly with position (dotted black line). The values are also slightly larger along the pink profile, which crosses the FGT/graphite/FGT heterostructure in region~1, particularly within the bottom FGT layer. This is consistent with the EELS measurements, which show that the lamella is slightly thicker in the bottom FGT layer in region~1 compared to region~2 (Fig.~S1).

A line profile showing the magnetic induction in the graphite and vacuum spacers is shown in Fig.~\ref{fig:Bfield}(e). The magnetic induction is similar on either side of the boundary between regions~1 and~2. Towards the left side of region~1, however, the amplitude of the fluctuations in $|\boldsymbol{B}_{\perp}(x,y)|\,t$ appears to decrease compared with that near the boundary. This may be related to the smaller domain width in the bottom FGT layer in this part of region~1.

In addition, the magnetic induction within the vacuum spacer increases as the interlayer separation decreases. From the line profile in Fig.~\ref{fig:Bfield}(e), we estimate the reduction in $|\boldsymbol{B}_{\perp}(x,y)|\,t$ within the vacuum spacer relative to the interior value extrapolated from the linear trend in the bulk of the FGT. At the first instance of interlayer domain misalignment, corresponding approximately to a separation of $\lambda$, the reduction is 23\%. At the largest spacer separation in region~2, the reduction increases to 52\%.

The average values of $|\boldsymbol{B}_{\perp}(x,y)|\,t$ within the upper and lower black boxes in Fig.~\ref{fig:Bfield}(c) are 30.9~T\,nm and 50.6~T\,nm, respectively. Using the EELS thickness measurements at the corresponding positions (Fig.~1), specifically points~8--11 for the top FGT layer and points~15--18 for the bottom FGT layer, we obtain $|\boldsymbol{B}_{\perp}| \approx 220$--$280$~mT for the top layer and $|\boldsymbol{B}_{\perp}| \approx 230$--$280$~mT for the bottom layer.

In summary, OAEH enables quantitative analysis of the magnetic induction and interlayer stray-field coupling in magnetic vdW heterostructures. In the interior of the FGT layers, we extract projected in-plane magnetic induction values of $\sim$220--280~mT. We further estimate that domain alignment between the top and bottom layers begins to break down when the field in the vacuum spacer is reduced by approximately 23\% relative to the bulk value, which occurs at separations on the order of $\lambda$. 

\subsection{Surface effects}

We next consider surface-related effects on the magnetic induction within the FGT layers. Figures~\ref{fig:Bfield}(c,\,d) show dome-shaped regions near the FGT surfaces, between magnetic domain walls, where $|\boldsymbol{B}_{\perp}(x,y)|\,t$ is reduced compared with the interior of the FGT. The striped white arc in Fig.~\ref{fig:Bfield}(c) highlights one such near-surface region with reduced projected in-plane induction. These regions extend up to $\sim$100~nm into the FGT. They are visible at the FGT/SiO$_2$, FGT/graphite, and FGT/vacuum interfaces, as seen from the line profiles in Fig.~\ref{fig:Bfield}(e). Thus, the nature of the interface or surface does not appear to have a significant influence on this surface effect. However, the reduction in magnetic induction is less pronounced at the right side of Fig.~\ref{fig:Bfield}(c), where the two FGT layers are in closer proximity and the stray-field decay within the vacuum spacer is smaller.

\subsubsection{Modelling of the magnetization distribution using model-based iterative reconstruction}

\begin{figure}[]
	\scalebox{\figurescale}{\includegraphics[width=1\linewidth]{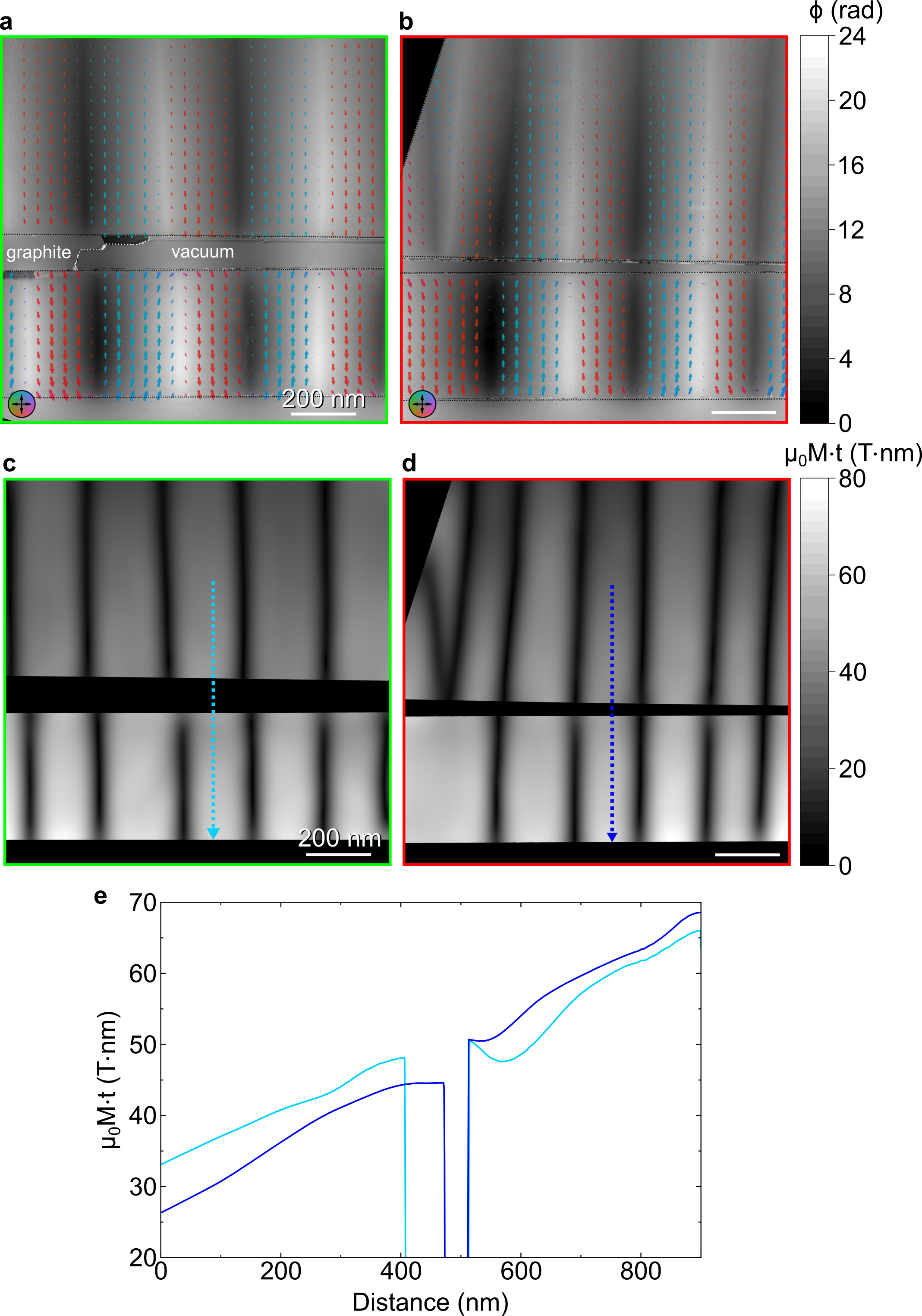}}
	\caption{\label{fig:MBIR}
        \textbf{Magnetization distributions obtained from a model-based iterative reconstruction.}
        \textbf{(a,b)} Magnetization distribution overlaid on the phase images used as input for the reconstruction. The thin black dashed lines indicate the boundaries of the specimen mask; see \textit{Methods} for details.
        \textbf{(c,d)} Maps of the magnitude of the magnetic polarization, $\mu_{0}|\boldsymbol{M}(x,y)|\,t$.
        \textbf{(e)} Line profiles extracted along the light- and dark-blue lines in (c) and (d), respectively.    
		} 
\end{figure}

To investigate the surface effect in greater detail, the specimen magnetization was reconstructed using a model-based iterative reconstruction approach with the experimental magnetic phase images as input \cite{caron2017model}. Further details are provided in \textit{Methods}. The reconstructed maps of the projected in-plane magnetization, $\boldsymbol{M}_{\perp}(x,y)$, obtained from the green and red boxes in Fig.~\ref{fig:Bfield}(a), are plotted as arrays of vectors in Fig.~\ref{fig:MBIR}(a,\,b). The maps are overlaid on the phase images used as input to the reconstruction algorithm. The magnitude of the magnetization integrated along the electron-beam direction, $|\boldsymbol{M}_{\perp}(x,y)|\,t$, is shown in Fig.~\ref{fig:MBIR}(c,\,d). Finally, line profiles of $|\boldsymbol{M}_{\perp}(x,y)|\,t$ taken along the light- and dark-blue lines in Fig.~\ref{fig:MBIR}(c,\,d), respectively, are shown in Fig.~\ref{fig:MBIR}(e).

We note that an apparent asymmetry in the projected in-plane induction between alternating domains in the top FGT layer is enhanced by the model-based reconstruction (Fig.~\ref{fig:Bfield}(d)). A similar asymmetry is also visible in the raw data; see Fig.~S10 for line profiles of $|\boldsymbol{B}_{\perp}(x,y)|\,t$ across the domains in Fig.~\ref{fig:Bfield}(d). At present, the origin of this asymmetry is not fully understood. It could, for example, arise from exchange bias associated with surface oxidation, including oxidation of the lamella sidewalls \cite{nogues1999exchange}, or from the stray field generated by the thicker edge of the lamella attached to the FIB lift-out grid. The latter possibility is illustrated and discussed in Fig.~S11.

As in the experimental data, the magnitude $|\boldsymbol{M}_{\perp}(x,y)|\,t$ is larger in the bottom FGT layer than in the top FGT layer because of the larger lamella thickness in the bottom layer. In addition, these results show that the dome-shaped regions of reduced projected in-plane magnetic induction observed in the experimental phase-derived maps (Fig.~\ref{fig:Bfield}(c,d)) are not reproduced in the reconstructed magnetization distribution. Notably, demagnetizing-field effects are not directly reflected in the magnetization distribution obtained by model-based iterative reconstruction. This suggests that the observed surface effect may arise from demagnetizing and stray fields, which we investigate further in the following section using micromagnetic simulations.

\begin{figure*}[]
	\scalebox{\figurescale}{\includegraphics[width=1\linewidth]{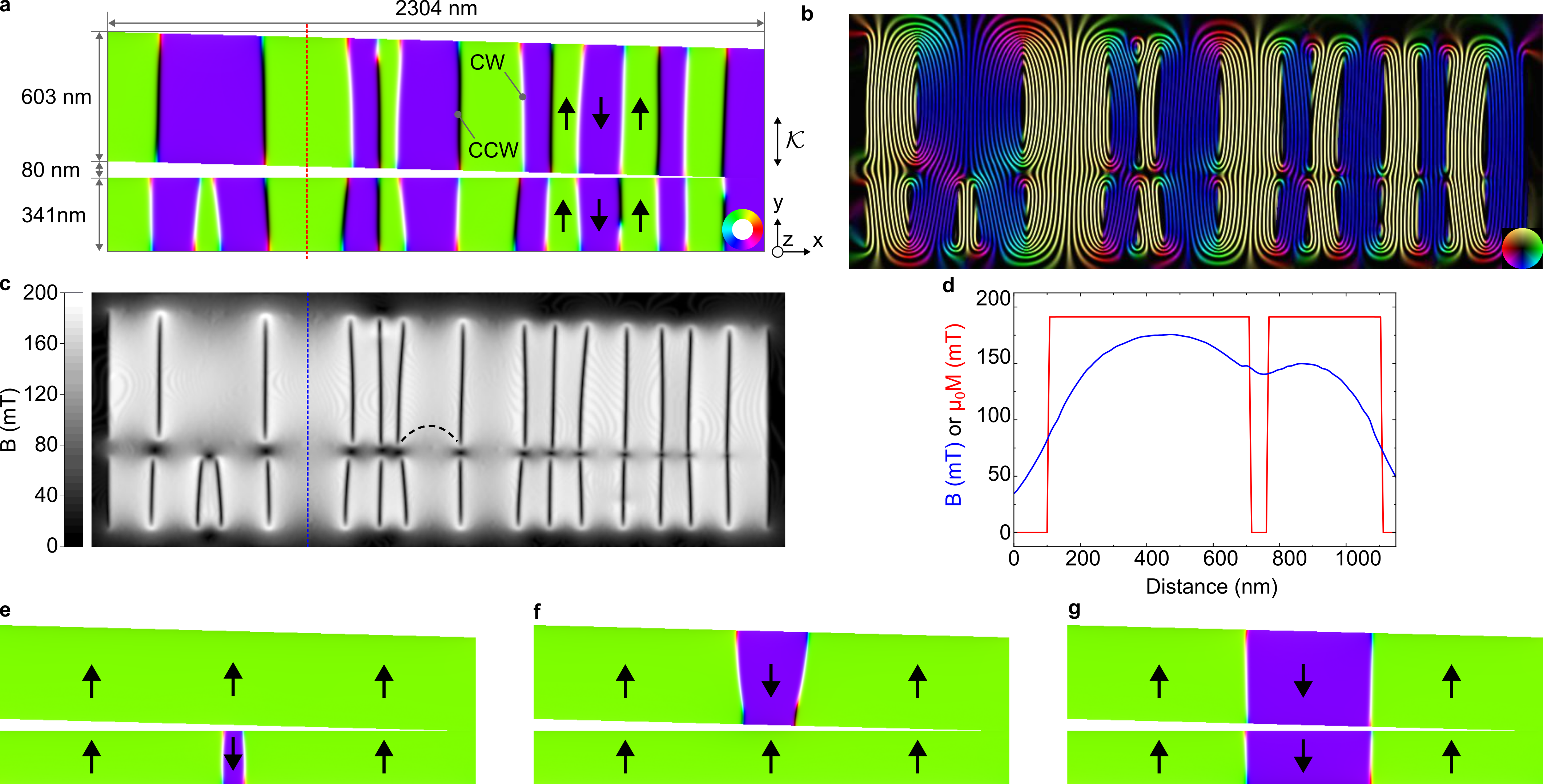}}
	\caption{\label{fig:simulation}
    \textbf{Micromagnetic simulations.}
    \textbf{(a)} The simulated sample consists of two 224~nm-thick stripes of different heights, stacked vertically and separated by a wedge-shaped vacuum spacer. The panel shows the magnetization configuration in the central plane of the sample after full energy minimization at zero external field, starting from a random initial magnetization. White and black domain walls correspond to clockwise (CW) and counterclockwise (CCW) Bloch-type domain walls, respectively.
    \textbf{(b)} Magnetic contour map calculated from the simulated phase image of the structure in \textbf{(a)}. The contour spacing is $2\pi/8$~rad.
    \textbf{(c)} Map of in-plane projected magnetic induction, $\mathbf{B}_{\perp}$, calculated from the simulated phase image.
    \textbf{(d)} Line profile of $\mu_{0}\mathbf{M}_{\perp}$ and $\mathbf{B}_{\perp}$ extracted along the red and blue dotted lines in \textbf{(a)} and \textbf{(c)}, respectively,
    \textbf{(e--g)} Equilibrium domain widths for three different configurations: domains present only in the thin (bottom) layer \textbf{(e)}, only in the thick (top) layer \textbf{(f)}, and simultaneously in both layers \textbf{(g)}.    
    }
\end{figure*}

\subsubsection{Magnetization and magnetic field distributions from micromagnetic simulations}

Micromagnetic simulations were performed for a sample with a geometry similar to that of HS1. The simulated structure (Fig.~\ref{fig:simulation}(a)) consists of two stripes approximately $2.3~\mu$m long along the $x$-direction and $224$~nm thick along the $z$-direction. The easy-axis anisotropy is oriented along the $y$-axis. The stripe width and length are indicated in the figure. In addition, a wedge-shaped vacuum spacer was introduced between the layers, with a maximum separation of 80~nm. The simulations show that a minimal micromagnetic model including exchange, uniaxial anisotropy, and magnetostatic interactions is sufficient to qualitatively reproduce the experimentally observed domain structure. For simulation details and material parameters, see \textit{Methods}.

Figure~\ref{fig:simulation}(a) shows one of the equilibrium configurations obtained by energy minimization from a random initial magnetization distribution. This energy-minimization procedure mimics the experimental ZFC process. On the right-hand side of the structure, where the vacuum spacer is thin, the domain walls in the top and bottom stripes are well aligned. In contrast, on the left-hand side, where the spacer thickness is largest, the domain wall positions in the two layers become misaligned, in qualitative agreement with the experimental results.

The corresponding magnetic contour map and induction map ($|\boldsymbol{B}_{\perp}(x,y)|$) are shown in Fig.~\ref{fig:simulation}(b,\,c), respectively. These are calculated from a simulated phase image of the equilibrium configuration shown in Fig.~\ref{fig:simulation}(a).

These maps qualitatively reproduce the experimental results and previous analysis. Specifically, the map of the projected in-plane magnetic induction (Fig.~\ref{fig:simulation}(c)) reproduces the main features of the experimental maps (Fig.~\ref{fig:Bfield}(c,\,d)), including the dome-shaped regions of reduced projected in-plane magnetic induction near the FGT surfaces. One such region is highlighted by a black dashed line.

Similarly, the magnetization profile from the simulation is consistent with the results obtained from the model-based iterative reconstruction (Fig.~\ref{fig:MBIR}(c,\,d)), with the magnitude of the projected in-plane magnetization remaining nearly constant throughout the sample, including near the FGT surfaces. A line profile of the magnitude of the projected in-plane magnetization, $\mu_{0}|\boldsymbol{M}_{\perp}|$, extracted along the red dotted line in Fig.~\ref{fig:simulation}(a), is shown in Fig.~\ref{fig:simulation}(d). The magnetization is extracted from the simulated structure by integrating along the electron-beam direction and projecting onto the image plane. For comparison, a line profile of $|\boldsymbol{B}_{\perp}(x,y)|$, extracted along the blue dotted line in Fig.~\ref{fig:simulation}(c), is also shown in Fig.~\ref{fig:simulation}(d). The profiles show a reduced magnitude of $\boldsymbol{B}_{\perp}$, particularly near the surfaces, compared with $\mu_{0}\boldsymbol{M}_{\perp}$. These results show that the experimentally observed magnetic induction distribution can be understood by the effects of the $\boldsymbol{H}$-field.

Finally, we investigate the effect of magnetic domain alignment on the domain width using micromagnetic simulations. Figures~\ref{fig:simulation}(e,\,f) show equilibrium configurations in which an isolated domain is present only in the bottom layer or only in the top layer, respectively. Figure~\ref{fig:simulation}(g) shows the case in which equilibrium-sized domains are aligned in both layers through magnetostatic coupling across the vacuum spacer. The domain width increases with layer thickness and reaches its maximum when domains in the top and bottom layers are magnetostatically coupled. This behavior is consistent with the trends summarized in Table~1.

In summary, HS1 exhibits regions of reduced projected in-plane magnetic induction in FGT extending up to $\sim$100~nm from the interfaces with graphite, SiO$_2$, and vacuum. Model-based iterative reconstruction of the magnetization using the OAEH data as input indicates that the surfaces do not strongly affect the magnitude or direction of the magnetization. This interpretation is corroborated by micromagnetic simulations, which show a qualitatively similar magnetization distribution as the model-based iterative reconstruction, while the corresponding magnetic induction maps reproduce the experimentally observed surface effects. We therefore conclude that these surface effects are primarily due to $\boldsymbol{H}$-fields--demagnetizing and stray fields--which reduce the magnetic induction near the FGT surfaces.

\subsection{Magnetic domain walls in FGT}

The domain wall topology in FGT remains an open question. In plan-view LTEM studies of vdW magnets, domain wall contrast that appears only upon tilting is often associated with N\'eel-type walls \cite{liu2023controllable, park2021neel, li2023visualizing}. However, similar behavior can also arise from narrow Bloch-type walls or multilayer Bloch walls with alternating chirality \cite{morikawa2015lorentz, savchenko2023magnetic, salikhov2025multilayer}. This question is closely related to DMI: bulk FGT is centrosymmetric and therefore is not expected to host intrinsic bulk DMI, although defects \cite{chakraborty2022magnetic}, non-stoichiometry \cite{liu2023controllable}, or surface oxidation \cite{park2021neel, peng2021tunable} may generate effective DMI.

We investigate the domain wall topology using plan-view LTEM, cross-sectional OAEH, and micromagnetic simulations; further details are provided in Supplementary Note 1 and Figs.~S12--S18. EDX measurements show that the FGT crystals are Fe-rich, with a composition of Fe$_{3.31}$GeTe$_{1.98}$ (Fig.~S12). In plan-view LTEM, domain wall contrast appears only upon tilting the sample (Fig.~S13), consistent with the behavior commonly associated with N\'eel-type walls. However, cross-sectional OAEH measurements show that the domain walls are very narrow, with a width of 8.9 $\pm$ 1.5~nm (Fig.~S14), close to the spatial resolution of AOEH (Fig.~S15). This is consistent with prior measurements of FGT domain wall widths using spin-polarized scanning tunneling microscopy and low-energy electron microscopy \cite{yang2022magnetic, tyson2025surface}, showing that FIB-preparation does not strongly affect the domain wall with in FGT. Such narrow Bloch-type walls could therefore produce similar tilt-dependent contrast.

Micromagnetic simulations reproduce the main experimental domain structures without including DMI and show that N\'eel- and Bloch-type walls in cross-sectional samples produce similar OAEH and LTEM signals (Figs.~S16, S17). The differences between these signals cannot be distinguished reliably at the present experimental noise level. We also investigated whether FeTe$_2$ surface layers, recently proposed as a possible source of DMI in annealed FGT \cite{xia2026spontaneous}, form under our heterostructure fabrication conditions. No such surface structure was observed in a graphite-capped FGT flake fabricated under conditions equivalent to those used for Heterostructures~1--4 (Fig.~S18).

We therefore conclude that, although the tilt-dependent LTEM contrast is consistent with N\'eel-type walls, it does not uniquely determine the domain wall topology. Furthermore, we find that LTEM and OAEH data from cross-sectional samples cannot distinguish between Bloch- and N\'eel-type walls at the current noise levels. Additional experiments, such as LTEM imaging under an applied in-plane magnetic field \cite{peng2021tunable}, would be required to distinguish Bloch- and N\'eel-type walls unambiguously.

\section{Conclusions and outlook}
We have quantified magnetic interactions between FGT layers in FGT/graphite/FGT vdW heterostructures with varying FGT separations by directly imaging the domain structure and measuring the magnetic field within and between the FGT layers. A characteristic stray-field coupling length scale of $\lambda = 37 \pm 7$~nm is extracted for the cross-sectional lamella geometry studied here, corresponding to the average separation at which the first misalignment between the layers is observed. In addition, magnetic induction measurements show that a separation on the order of $\lambda$ corresponds to an approximately 23\% reduction in the field within the spacer relative to that in the interior of the FGT. Together, these results connect a directly observed characteristic coupling length ($\lambda$) to a quantitative reduction in the local magnetic field within the vacuum spacer and demonstrate that the effective interaction strength between vertically stacked FGT layers can be tuned from strong to weak coupling through the interlayer separation and sample geometry.

Furthermore, surface effects lead to a reduction of the in-plane projected magnetic induction over length scales of up to $\sim$100~nm from an FGT surface. By comparing experimental data with model-based iterative reconstruction of the magnetization and micromagnetic simulations, we determine that these surface effects are due to demagnetizing and stray fields. These findings suggest that surface effects can dominate the local magnetic induction in flakes with thicknesses $\lesssim$100~nm.

Considering the magnetic domain walls in FGT (Fe$_{3.31}$GeTe$_{1.98}$) we find tilt dependent LTEM contrast in plan-view flakes suggesting N\'eel-type domain walls. Nevertheless, we emphasize that current noise levels in both LTEM and OAEH cannot unambiguously determine the domain wall-type, requiring additional methods, such as application of in plane magnetic fields to plan-view samples \cite{peng2021tunable}.

Overall, these results provide quantitative insight into interlayer magnetic coupling and surface effects in magnetic vdW heterostructures, while demonstrating an approach that can be extended from the bulk-like thickness regime studied in this work toward more device-relevant structures. We outline some practical limitations of the techniques. First, reliable determination of the magnetic induction in an FGT flake imaged in cross-section requires the FGT thickness to be comparable to or greater than the spatial resolution of OAEH, which in our case is on the order of 10~nm. Higher spatial resolution may be achievable using phase-shifting electron holography \cite{yamamoto2021phase} or a TEM equipped with an objective lens that can be operated under magnetic-field-free conditions, potentially enabling the study of FGT flakes thinner than 10~nm. Second, because LTEM is an out-of-focus imaging technique, Fresnel diffraction from sample edges can hinder the visualization of domain walls in cross-sectional specimens. We estimate that FGT thicknesses of approximately 20~nm or greater are required to confidently visualize domain walls using LTEM in cross-sectional specimens; Fig.~S4(c), for example, shows LTEM images of a 42~nm-thick FGT flake. Regarding the lamella thickness along the electron-beam direction, both techniques have similar constraints. A maximum lamella thickness of approximately 300~nm is required to maintain sufficient electron transparency, whereas the minimum practical thickness is more likely to be limited by FIB preparation than by imaging or signal constraints.

We anticipate that OAEH studies of cross-sectional specimens can be extended to probe proximity interactions between vdW ferromagnets and a wide range of other vdW materials. Such studies could include heterostructures incorporating heavy-element materials that induce strong interfacial spin–orbit coupling, as well as superconductors, to investigate interactions between magnetic domains and flux vortices \cite{diaz2024manipulating}. Other techniques, such as vector field electron holography, can reconstruct the three-dimensional magnetic field in these structures and provide additional information on stray fields and interface effects \cite{vsilinga2025model}. Finally, the response of such heterostructures to external stimuli, including temperature variations and applied magnetic fields, could be directly visualized.

\section{Methods}

\footnotesize{

\subsection{Growth of bulk Fe$_3$GeTe$_2$ crystals}
Bulk Fe$_3$GeTe$_2$ crystals were prepared by chemical vapor transport (CVT) in sealed quartz ampoules using chlorine as the transport agent. Fe (99.9\%, $-100$ mesh, Strem, USA), Ge (99.999\%, $-100$ mesh, Wuhan Xinrong New Materials Co., China), and Te (99.999\%, $-100$ mesh, Wuhan Xinrong New Materials Co., China) were mixed in a stoichiometric ratio corresponding to 15~g of Fe$_3$GeTe$_2$. TeCl$_4$ (99.9\%, Strem, USA; 0.6~g) was added as the transport reagent, and the mixture was sealed in a quartz ampoule under high vacuum ($<1\times10^{-3}$~Pa) using an oil diffusion pump with a liquid-nitrogen cold trap.

The ampoule was placed in a muffle furnace and gradually heated to 750\,$^{\circ}$C for 50~h. The resulting polycrystalline Fe$_3$GeTe$_2$ was then homogenized in the sealed ampoule and transferred to a two-zone horizontal furnace. The growth zone was held at 800\,$^{\circ}$C and the source zone at 700\,$^{\circ}$C for 2~days. Subsequently, the growth and source zones were set to 750\,$^{\circ}$C and 650\,$^{\circ}$C, respectively, for 14~days. Finally, the furnace was cooled to room temperature and crystals were collected in an argon-filled glovebox.

\subsection{vdW heteostructure assembly}
Graphite and FGT flakes were obtained by mechanical exfoliation of natural graphite (NaturGrafit GmbH, Germany) and bulk Fe$_3$GeTe$_2$ crystals, respectively, onto Si substrates with 90~nm-thick SiO$_2$ using 3M Magic Scotch tape. Heterostructures were assembled using a dry-transfer method based on a polycarbonate (PC)-coated polydimethylsiloxane (PDMS) stamp mounted on a glass slide for optical microscopy. Exfoliation and heterostructure assembly were performed in a glovebox under an inert N$_2$ atmosphere. The highest temperature reached during the dry-transfer method is 190\,$^{\circ}$C, when the PC is melted to release the final heterostructure onto the SiO$_2$ surface.

\subsection{FIB preparation}
Cross-sectional TEM lamellae were prepared from as-assembled FGT/graphite/FGT heterostructures on Si/SiO$_2$ substrates using Ga$^{+}$ focused ion beam milling in an FEI Helios dual-beam FIB-SEM operated at 30~kV. A carbon protection layer was deposited before milling. The ion beam current was successively reduced as the lamella thickness decreased starting from 0.77~nA down to 83~pA. The final thinning on the FIB lift out grid was performed at 2~kV to reduce surface damage. We estimate that the lamellae were thinned approximately 5~nm in this step. For HS1, an additional protection layer thinning step was performed at 5~kV by milling the protection layer in a direction parallel to the FGT crystal \textit{c}-axis. During this step, the lamella thickness was further reduced by approximately 150~nm as estimated by SEM images (Fig.~S1). See Figs.~S1-S4 for additional details about HS1-HS4, respectively. 

\subsection{Plan-view sample preparation}
FGT flakes were obtained by mechanical exfoliation of bulk Fe$_3$GeTe$_2$ crystals onto Si substrates with 90~nm-thick SiO$_2$ using 3M Magic Scotch tape. Suitable flakes were identified by optical microscopy. Exfoliation and optical inspection were performed in an N$_2$-filled glovebox. The flakes were then transferred to TEM grids (Norcada Inc., Canada) outside the glovebox using cellulose acetate butyrate as a polymer handle \cite{schneider2010wedging}. Finally, the polymer film was dissolved in acetone, and the grids were rinsed in isopropanol and dried in air.

\subsection{Atomic force microscopy}
AFM measurements of the thickness of as-exfoliated flakes were conducted using a Bruker Dimension Edge AFM operated in tapping mode. 

\subsection{EELS measurements}
EELS data were acquired using a Thermo Fisher Scientific Titan G2 80-200 ChemiSTEM operated at 200~kV and equipped with a Gatan Enfinium ER 977 post-column energy filter. EELS measurements provide the specimen thickness in units of the inelastic mean free path, $\lambda_{\mathrm{EELS}}$, i.e., $t/\lambda_{\mathrm{EELS}}$. To calibrate $\lambda_{\mathrm{EELS}}$ for FGT, three exfoliated plan-view FGT flakes were transferred onto TEM grids (Fig.~S5(a,b)). Their thicknesses were independently measured using atomic force microscopy (Fig.~S5(c--h)), after which EEL spectra were acquired from the same flakes (Fig.~S5(i)). The FGT thicknesses measured by atomic force microscopy are plotted against the corresponding EELS measurements, $t/\lambda_{\mathrm{EELS}}$, in Fig.~S5(j). From this calibration, we obtain $\lambda_{\mathrm{EELS}} = 128$~nm.

\subsection{Lorentz TEM and off-axis electron holography}
LTEM and OAEH were performed in an image-$C_S$-corrected Thermo Fisher Scientific Titan 80-300 microscope equipped with two electron biprisms and operated at 300~keV. A liquid-nitrogen-cooled double-tilt specimen holder (Gatan model 636) was used to vary the specimen temperature between 95 and 300~K. LTEM images and off-axis electron holograms were recorded under magnetic-field-free conditions using a direct electron-counting camera (Gatan K2 IS). Phase images were reconstructed from holograms using standard fast Fourier transform algorithms (Holoworks 6 plugin in Gatan DigitalMicrograph).

The electron phase shift contains both magnetic and electrostatic contributions. To separate these, holograms of the same region were acquired at 95~K and at room temperature (above $T_{\mathrm{c}}$ of FGT), and the resulting room temperature phase map was subtracted from the low temperature one. Magnetic contour and phase shift maps were generated in DigitalMicrograph, and projected in-plane induction maps were computed from the phase images using Eq.~\ref{eq:B} with a custom Python script.

\subsection{Model-based iterative reconstruction}

The projected magnetization distribution was reconstructed from the magnetic contribution to the electron-optical phase using the model-based iterative reconstruction approach and the code outlined in \cite{caron2017model}. In this approach, a forward model is used to calculate the magnetic phase expected from a trial magnetization distribution. The calculated phase is compared with the experimentally measured magnetic phase, and the magnetization distribution is adjusted iteratively to minimize their difference. First-order Tikhonov regularization is used to suppress unrealistically rapid spatial variations in the reconstructed magnetization and to stabilize the reconstruction against experimental noise. 
Prior knowledge of the specimen geometry is incorporated through a mask, which defines the region in which the projected magnetization is allowed and excludes non-magnetic regions from the reconstruction. Finally, since the model assumes a homogeneous sample thickness, a constant thickness of 166 nm was assumed for the reconstruction. The resulting reconstructed magnetization is then multiplied by this thickness to yield maps of $|\boldsymbol{\mu_{0}M}_{\perp}(x,y)|\,t$. Dividing by the true local lamella thickness will allow extracting $|\boldsymbol{\mu_{0}M}_{\perp}(x,y)|$.

\subsection{Micromagnetic simulations}
In our simulations, we use the following micromagnetic energy functional, which includes the exchange energy, uniaxial anisotropy with the easy axis oriented along the $y$ direction, and the energy of the demagnetizing fields~\cite{Fratta}:
\begin{align}
\mathcal{E}=
\int\limits_{V_\mathrm{m}}&\left[
\mathcal{A}\sum_{i=x,y,z} |\nabla m_i|^2
-M_\mathrm{s}\,\mathbf{m}\cdot(\nabla\times\mathbf{A})
-\mathcal{K}\,(\mathbf{m}\cdot\mathbf{e}_\mathrm{y})^2\right]\!\mathrm{d}\mathbf{r}
\nonumber\\
&+ \frac{1}{2\mu_0}
\int\limits_{\mathbb{R}^3}\left[
\sum_{i=x,y,z} |\nabla A_{i}|^2\right]\!\mathrm{d}\mathbf{r},
\label{Ham_m}
\end{align}
where $\mathbf{m}(\mathbf{r})=\mathbf{M}(\mathbf{r})/M_\mathrm{s}$ is the unit magnetization field, $M_\mathrm{s}=|\mathbf{M}(\mathbf{r})|$ is the saturation magnetization, and $\mu_0$ is the vacuum permeability.
The parameters $\mathcal{A}$ and $\mathcal{K}$ denote the exchange stiffness and the uniaxial anisotropy constant, respectively.
The vector field $\mathbf{A}(\mathbf{r})$ in Eq.~\eqref{Ham_m} represents the magnetic vector potential generated by the magnetization of the sample.
The first integral in Eq.~\eqref{Ham_m} is taken over the volume of the magnetic sample $V_{\mathrm{m}}$, while the second integral is evaluated over the entire space $\mathbb{R}^3$ to account for the demagnetizing field.
All calculations were performed using the \textsc{Excalibur} micromagnetic code~\cite{Excalibur}; implementation details are described in Ref.~\cite{Zheng_21}.
For the simulations shown in Fig.~\ref{fig:simulation}, we used the following material parameters: $\mathcal{A}=3$~pJ/m, $\mathcal{K}=0.1$~MJ/m$^{3}$, and $M_\mathrm{s}=150$~kA/m.
The simulation domain has dimensions of $2304~\mathrm{nm}\times1024~\mathrm{nm}\times224~\mathrm{nm}$ along the $x$, $y$, and $z$ directions, respectively, and is discretized into $768\times256\times64$ cuboidal cells.

The material parameters were selected to provide the best overall agreement between the simulations and the experimentally observed magnetic configurations.
In particular, the chosen parameter set provides good quantitative agreement with the characteristic domain size and domain-wall width observed experimentally.
It also reproduces the experimentally measured Lorentz TEM contrast and electron holography signal.
The simultaneous agreement with these independent experimental observables supports the physical relevance of the material parameters used in the simulations.

\subsection{Néel versus Bloch type domain walls}
The micromagnetic simulations of the stripe domains with Néel and Bloch-type domain walls presented in Fig.~S16 were performed on a domain of size 1024~nm~$\times~512$~nm $\times~224$~nm, 
along the $x$, $y$, and $z$ directions, respectively. 
For these calculations, we also apply periodic boundary conditions along the $x$-axis. 
For domains with Bloch-type domain walls, Fig.~S13(a-d), we minimize the energy using the same material parameters as described in the above section.
The initial configuration in this simulation represents alternate domains with magnetization pointing along $+\mathbf{e}_z$ and $-\mathbf{e}_z$.
To stabilize the configuration with Néel-type domain wall, Fig.~S16(e-h), we added to Eq.~\eqref{Ham_m} an additional DMI term:
\begin{align}
\mathcal{E}_{\rm DMI}=
\int\limits_{V_\mathrm{m}}\mathcal{D}\left[
 m_z\frac{\partial m_x}{\partial r_x}- m_x\frac{\partial m_z}{\partial r_x}
\right]\!\mathrm{d}\mathbf{r}.
\label{DMI}
\end{align}
To ensure that the presence of DMI does not significantly affect the width of the domain wall, but only changes the magnetization rotation within the domain wall, we set the DMI constant to a small value, $\mathcal{D}=0.3$ mJ/m$^2$. 

\section{Acknowledgements}
This work was supported by the European Union’s Horizon 2020 Research and Innovation Programme (grant 856538, project 3D MAGIC). R.E.D.-B.\ acknowledges support from the European Union under grant agreement no. 101094299 (project IMPRESS). Z.S. acknowledges support by project LUAUS25268 from the Ministry of Education Youth and Sports (MEYS) and by the project Advanced Functional Nanorobots (reg.\ No.\ CZ.02.1.01/0.0/0.0/15\_003/0000444 financed by the EFRR). The authors gratefully acknowledge support from L.\ Risters and L.\ Kibkalo on FIB-preparation of the specimens. 
The authors would like to acknowledge C.\ Stampfer and B.\ Beschoten (RWTH Aachen) for access to their custom-built van der Waals transfer system for heterostructure assembly.

\bibliographystyle{naturemag-3of6}
\bibliography{references}

\end{document}